\documentclass[a4paper,10pt,twoside]{cpc-hepnp}

\usepackage{multicol}
\usepackage{graphicx}
\usepackage{booktabs}
\usepackage{amssymb,bm,mathrsfs,bbm,amscd}
\usepackage[tbtags]{amsmath}
\usepackage{lastpage}

\begin{document}

\fancyhead[c]{\small Submitted to Chinese Physics C~~~Vol. XX, No. X (XXX)
XXXXXX} \fancyfoot[C]{\small XXXXXX-\thepage}

\footnotetext[0]{Received 27 September 2013}

\title{Magnetic shield of PMT used in DAMPE electromagnetic calorimeter\thanks{Supported by the 973 Program (Grant No. 2010CB833002), and the Strategic Priority Research Program on Space Science of the Chinese Academy of Science (Grant No. XDA04040202-4) }}

\author{%
      WANG Pei-Long$^{1)}$\email{wplong@mail.ustc.edu.cn}%
\quad ZHANG Yun-Long$^{}$
\quad WANG Xiao-Lian$^{}$
\quad XU Zi-Zong$^{2)}$\email{zzxu@mail.ustc.edu.cn}%
}
\maketitle

\address{
State Key Laboratory of Particle Detection and Electronics, University of Science and Technology of China, Hefei 230026, China\\
}

\begin{abstract}
The magnetic characteristics of photomultiplier tube R5610A-01 are studied in this paper. The experimental data shows that the gain of R5610A-01 loses about 53\% when the magnetic field is 3G along its +X axis. A cylinder of one-layer permalloy strip is able to reduce the effect of 3G magnetic field on the PMT's gain to less than 1\%.
\end{abstract}

\begin{keyword}
photomultiplier tube, magnetic shield, characteristics
\end{keyword}

\begin{pacs}
07.55.Nk, 29.40.Mc
\end{pacs}

\footnotetext[0]{\hspace*{-3mm}\raisebox{0.3ex}{$\scriptstyle\copyright$}2013
Chinese Physical Society and the Institute of High Energy Physics
of the Chinese Academy of Sciences and the Institute
of Modern Physics of the Chinese Academy of Sciences and IOP Publishing Ltd}%

\begin{multicols}{2}

\section{Introduction}
DAMPE, DArk Matter Particle Explorer, is an equipment focusing on high energy electron and gamma ray detection. The program is building a 60$\times$60$\times$60cm BGO calorimeter which will be loaded on a satellite to study the properties of high energy particles in space. This BGO calorimeter will use 616 Hamamatsu R5610A-01 photomultiplier tubes (PMT). Because of the satellite's inner magnetic field and the earth's magnetic field, the maximum magnitude of the magnetic field that the BGO calorimeter will sustain is 2.1G. Since the PMT is very sensitive to magnetism \cite{lab1,lab2}, it is necessary for us to shield the PMT from the magnetic field. The simulation result from Geant4 shows that if the energy resolution of the BGO calorimeter for $e/\gamma$ wants to reach 1.5\% at 800GeV \cite{lab3}, then the PMT's signal variation affected by the magnetic field must be limited to less than $\pm$1\%. Very strict mass limitation of the satellite's load requires that the material for the magnetic shield must be as light as possible. Following the short introduction, the next sections will be: 2 Experimental setup; 3 Results and discussion; and 4 Conclusions.

\section{Experimental setup}
\subsection{Building a variable magnetic field}
To produce a stable and uniform magnetic field for the test of the PMT's magnetic characteristics, a Helmholtz coil was designed and assembled. The formula of the magnetic induction at the midpoint between two coils was used \cite{lab4}:
\begin{equation*}
B = \left (\frac{4}{5} \right)^{3/2} \frac{\mu_0 n I}{R},
\end{equation*}
where $\mu_0$ is the permeability of free space, n is the number of turns in each coil, I is the current flowing through the coils and R is the radius of the coils and also the distance between two coils.

To produce a magnetic field of more than 2.1G, the parameters of the Helmholtz coil were designed as follows:
\begin{center}
\tabcaption{ \label{tab1}  Design parameters of the Helmholtz coil}
\footnotesize
\begin{tabular*}{80mm}{c@{\extracolsep{\fill}}ccc}
\toprule Parameter & Value \\
\hline
Diameter of the coils &300mm \\
Number of turns in each coil & 18 \\
Current range flowing through the coils & $0\sim3.2A$\\
Magnetic field range & $0\sim3.77G$\\

\bottomrule
\end{tabular*}
\end{center}

The diameter of the coils was designed as 300mm. This can insure the field's uniform zone (about 100$\times$100$\times$100mm) is large enough for the location of PMT R5610A-01 ($\Phi$18.5$\times$30mm). The material of the coils' frame is nylon to avoid disturbing the coils' field.

As shown in Fig.1 the Z axis of R5610A-01 is located in the middle plane of the coils and parallel to the tube axis. The R5610A-01 can rotate around its Z axis to find the X and Y axis relative to the field's lines. A magnetometer with an accuracy of 0.001G was used to measure the magnetic field of the coils. The power supply that drove the Helmholtz coil has a current stability of 0.2\%.

\begin{center}
\includegraphics[width=8.2cm]{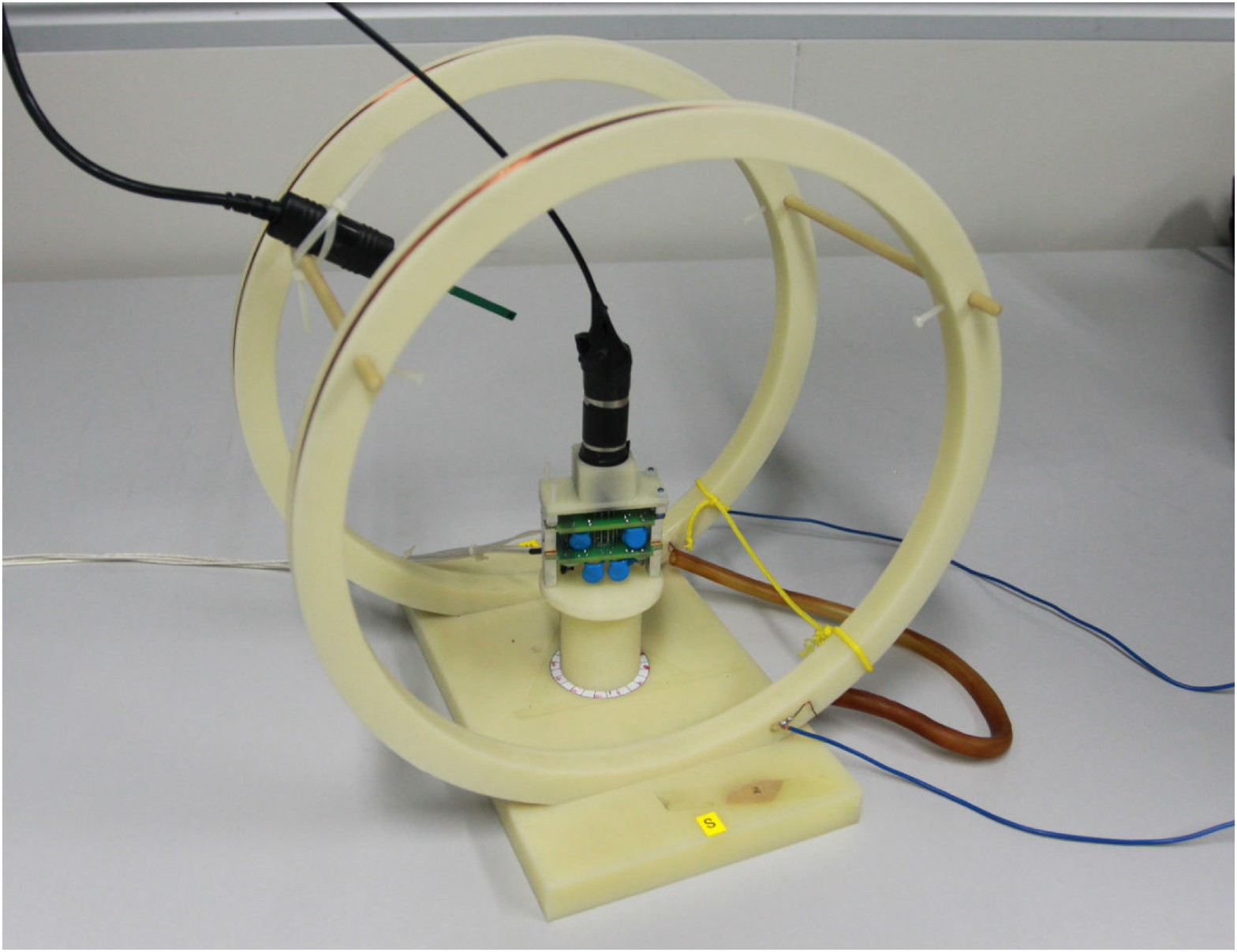}
\figcaption{\label{fig1}   The Helmholtz Coil. LED light was transmitted to PMT by optical fiber. The magnetometer sensor was fixed on the coil to monitor the magnetic field at the midpoint of the coils}
\end{center}

\subsection{Data acquisition system}
Fig.2 shows the block diagram of the data acquisition system. The blue LED was located far from the effects of the Helmholtz coil's field. An optical fiber guided light pulses to the entrance window of R5610A which sat on the rotating table between the coils. A pulse generator of the voltage stability better than 1\% drove the LED and synchronously sent a trigger pulse to the gate generator. CAEN N1470 with the voltage stability of $\pm$0.02\% in one week was used to supply the R5610A. The charge signal of dynode 8 responding to the LED's light pulse was sent to the FEE (Front-End Electronics), by which the charge signal was amplified and timed the gate pulse with the amplified voltage pulse. Then the digitized counts proportional to the charge would be output from an ADC chip and data were recorded by the computer. A Gaussian was fit to the spectra and the peak channel (ADC channel) was obtained.

\begin{center}
\includegraphics[width=8.4cm]{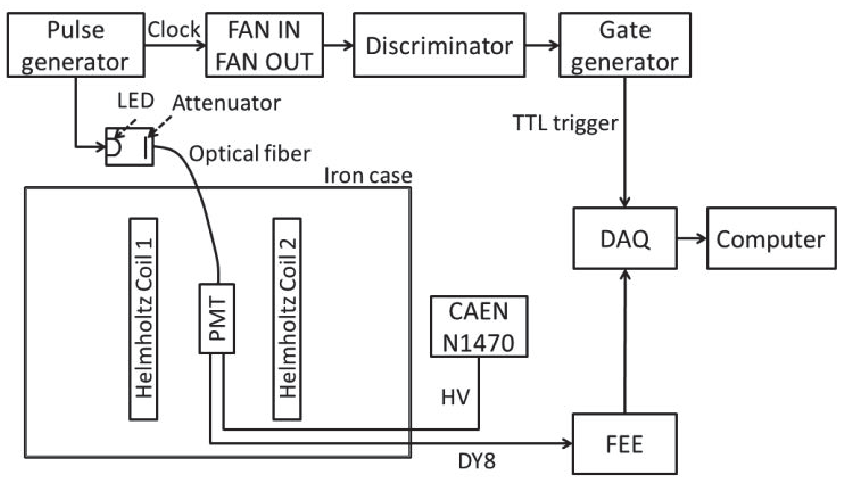}
\figcaption{\label{fig2}   Block diagram of the data acquisition system}
\end{center}

\subsection{Checking the stability of the whole system}
After building the whole test system, its stability should be checked. We adjusted the Helmholtz coil's field to cancel out the earth's magnetic field and then traced the PMT's signal variation in 0G for 24 hours with a constant light pulse driving. The results show that the period from 0:00 to 9:00am is the quietest time in 24 hours. Fig. 3 is the results acquired in the quietest period. ADC channel shows a normal distribution of the mean value of 4037.57 and $\sigma$ of 4.25. The stability is $\sim$0.1\% in 9 hours.

\begin{flushright}
\includegraphics[width=9.2cm]{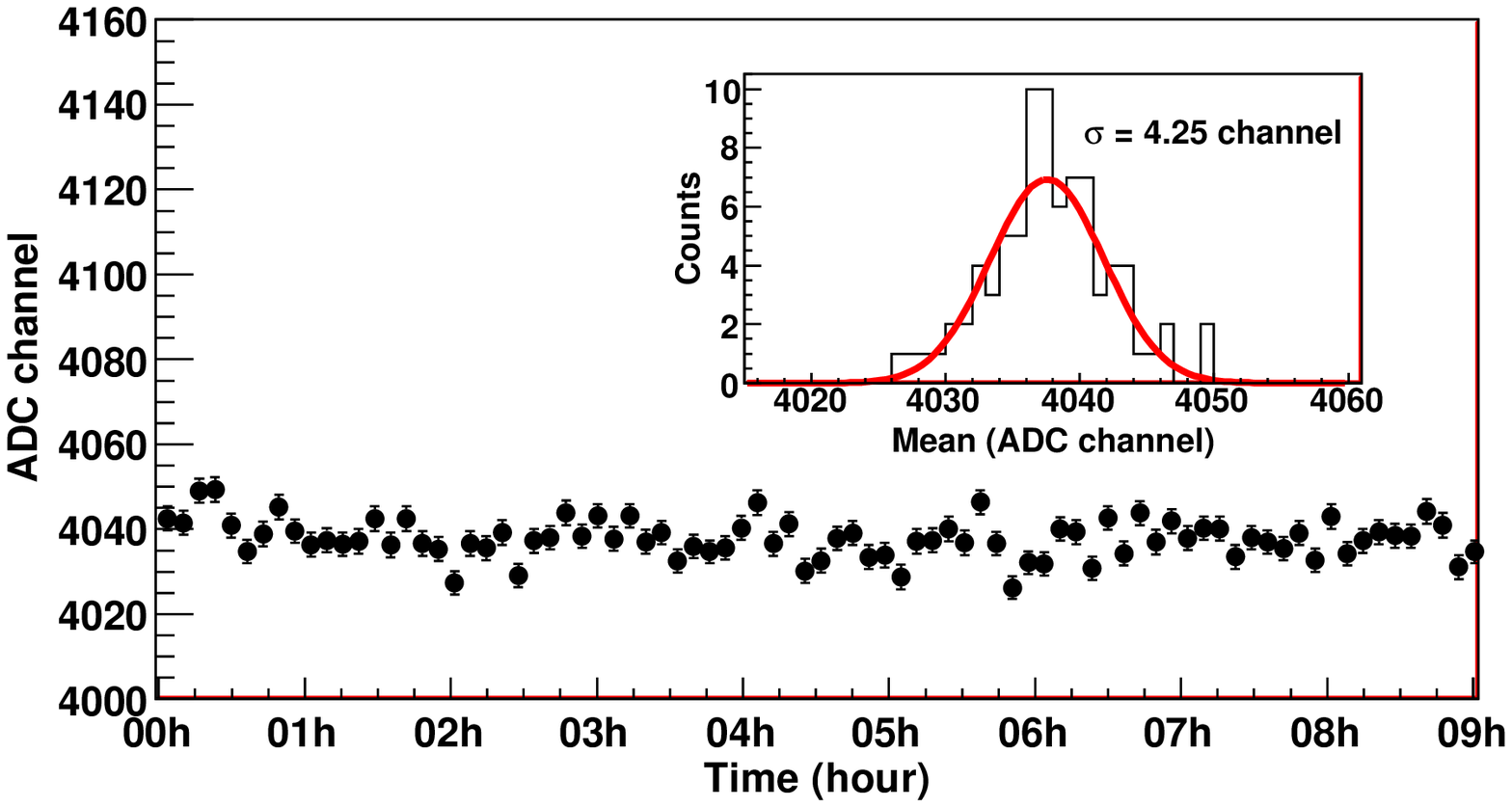}
\figcaption{\label{fig3}   Stability of the whole system from 0:00am to 9:00am in 0G}
\end{flushright}

\section{Results and discussion}
\subsection{PMT's most sensitive direction in X-Y plane}
In the photoelectron multiplication process of the PMT, a special configuration of the electric field is designed using a series of dynodes to accelerate, multiply and focus the electrons. The external magnetic field will destroy the electron paths and make the electrons evaded from the collection and multiplication of the next dynode. The first few dynodes are the most sensitive to the magnetic field because the electron kinetic energies are lower there and those electrons contribute to the final output with more weight. Due to the orientations of the first few dynodes, when the external magnetic field lines are perpendicular to electric field lines between dynodes the path destruction and the gain suppression is the greatest. The orientation of PMT which is most sensitive to the magnetic field will be defined as the X axis. To find this orientation, a 1G field of the Helmholtz coil was set while rotating the R5610A around Z axis (see Fig.1). The spectra of R5610A responding to a constant LED pulse with various rotating angles was taken. Fig.4 shows that the angle orientation of 270$^{\circ}$, where the gain suppression is the greatest, is the X axis, and the angle orientation of 180$^{\circ}$, where the gain suppression is the least, is the Y axis.
\begin{center}
\includegraphics[width=9.2cm]{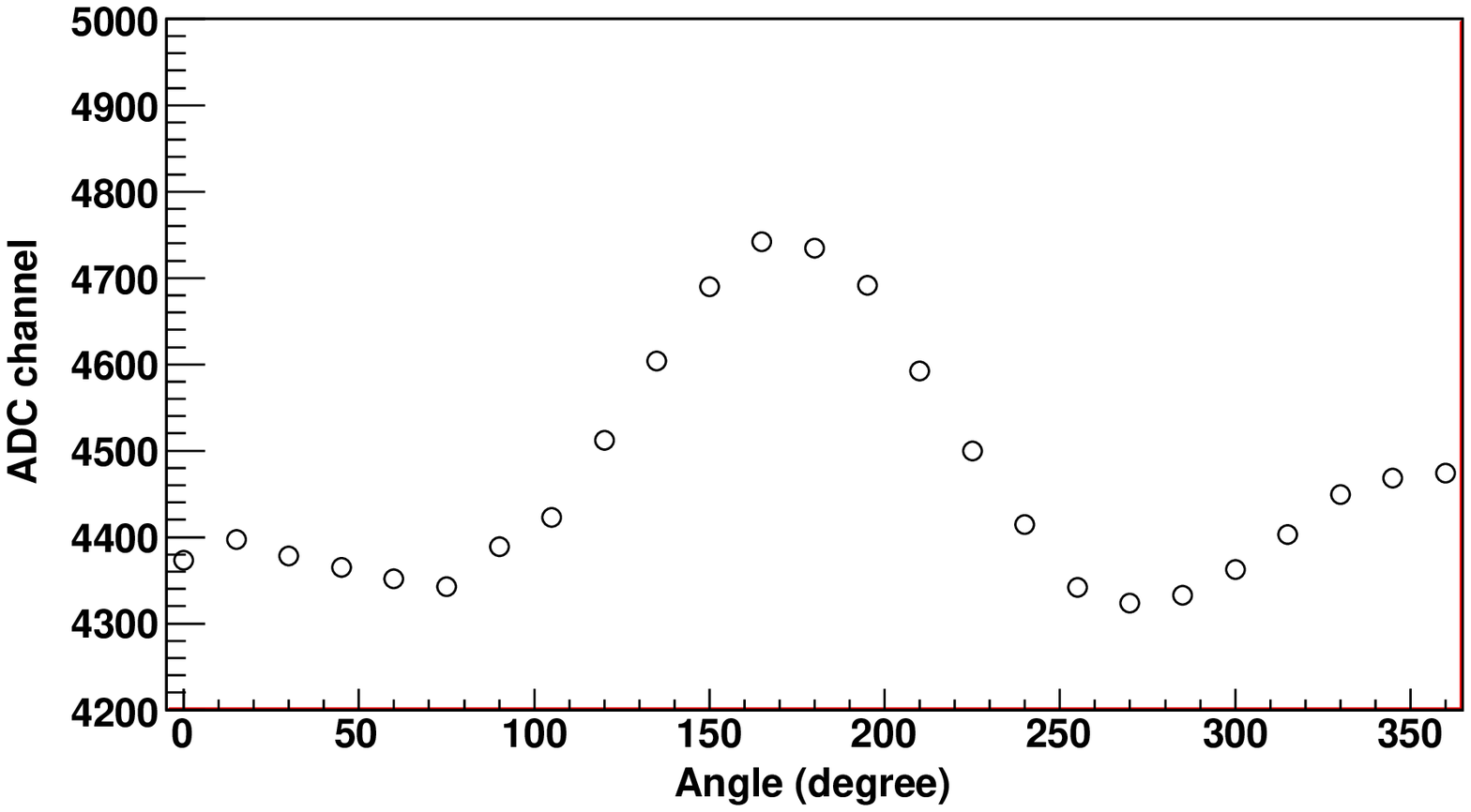}
\figcaption{\label{fig4}   The signal variation of the PMT with the angle of the magnetic field in X-Y plane when the magnetic magnitude is 1G.}
\end{center}

\subsection{Gain of the R5610A varies with the magnitude of the magnetic field}
Rotating the R5610A's X axis parallel to the field lines of Helmholtz coil, with constant LED light pulses and HV-setting, the spectra of the charge output from DY8 was taken under the field magnitude of 0, 1, 2 and 3G respectively. The data represented by full squares in Fig.5 show the peak channels vary with the magnetic field along +X (top) and -X (bottom) axis. Rotating the $\pm$Y axis and $\pm$Z axis of the R5610A along the direction of magnetic field, and repeating the same cycle as the $\pm$X case resulted in the data represented by full triangles and full circles in Fig.5, which show the peak channels vary with magnetic field along +Y and +Z (top), -Y and -Z (bottom) respectively.

As shown in Fig.5, the field of 3G in the $\pm$X direction of the R5610A makes its gain lose $\sim$53\% (+X) and 43\% (-X) compared to 0G-case. For the Y direction, the gain is much less sensitive to the magnetic field; gain loss is only 4\% ($\pm$Y). Up to a maximum 3G in the Z-direction there is no visible effect on the gain of the R5610A.
\begin{center}
\includegraphics[width=9.2cm]{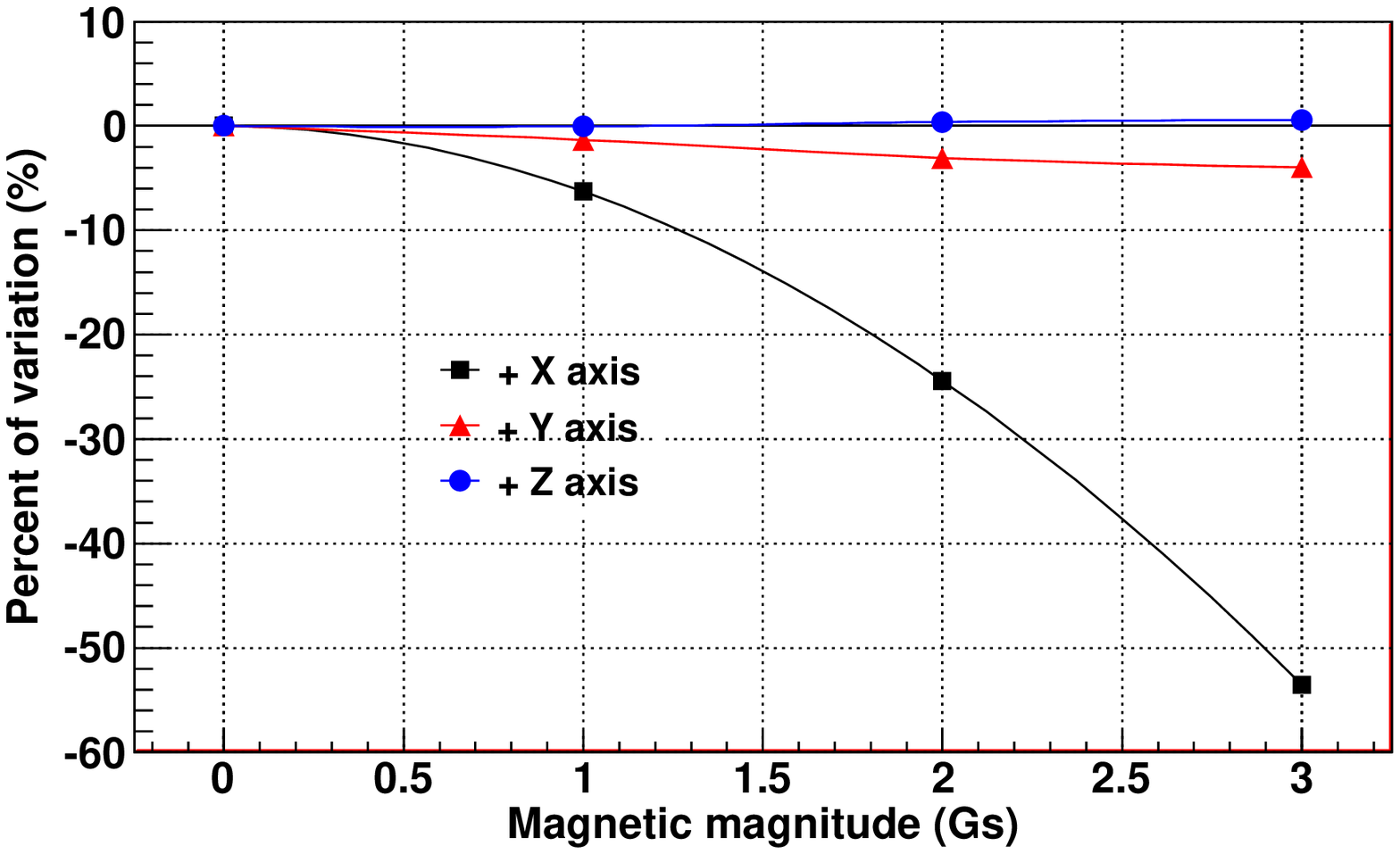}
\includegraphics[width=9.2cm]{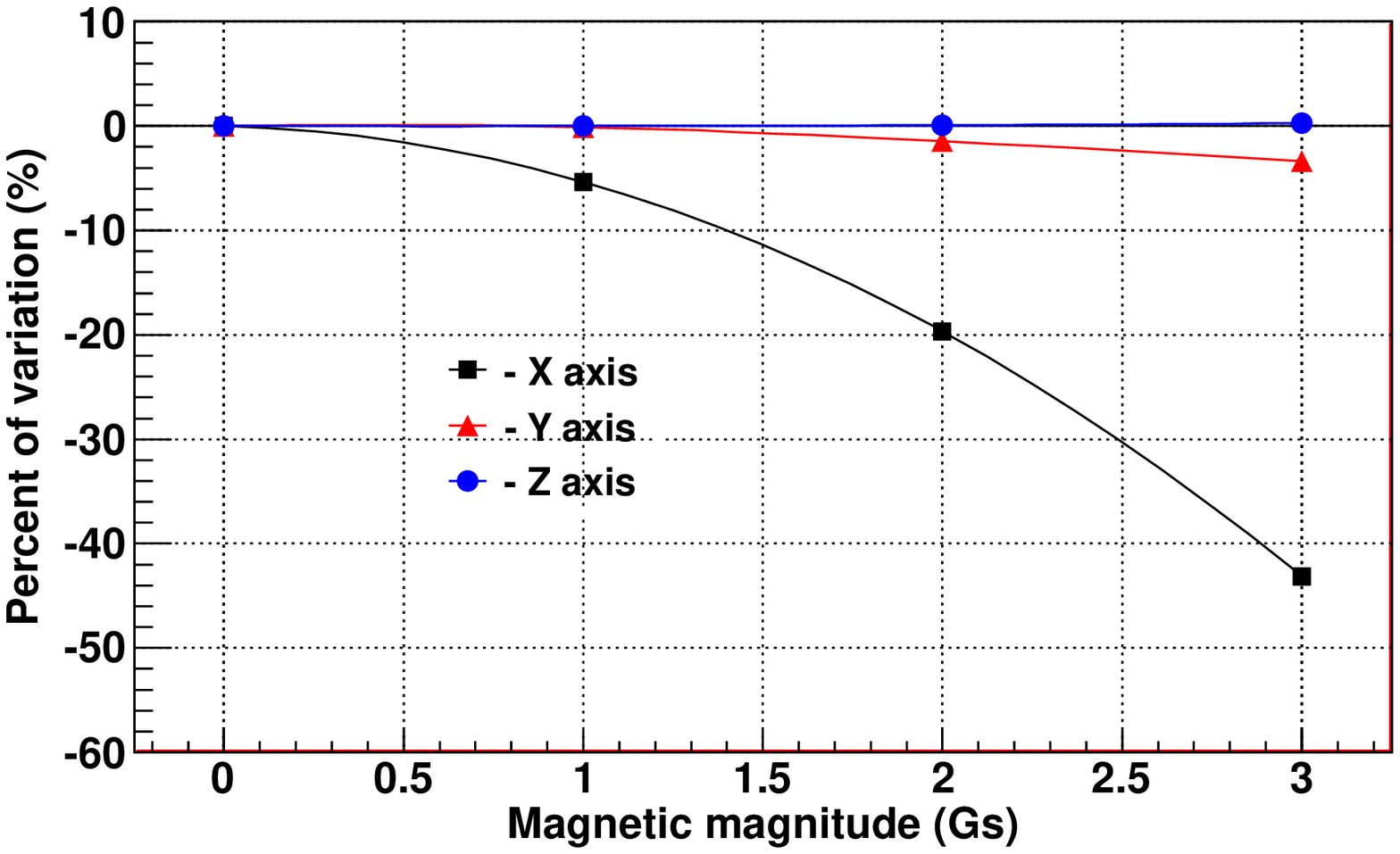}
\figcaption{\label{fig5}   The PMT's signal variation with the magnitude of magnetic field in $\pm$X, $\pm$Y and $\pm$Z directions.}
\end{center}

\subsection{Magnetic shield on PMT R5610A-01}
From the theory about the electromagnetic field \cite{lab5}, a film of thickness t and magnetic permeability $\mu_I$ wrapped as a closed cylinder of a radius r, can reduce the magnetic field from $H_{out}$ to $H_{in}$ which are perpendicular to the axis of the cylinder:
\begin{equation*}
\frac{H_{out}}{H_{in}}=(\frac{3t\mu_I}{4r})^n
\end{equation*}
n is the number of individual layers of the film to form the shield cylinder.

The permalloy strip (a kind of Fe-Ni-alloy of $\mu_{I}$=8000) of 27$\mu$m thickness and 25mm width was used as the shield cylinder with the inner diameter $\Phi$19mm matched with the R5610A's diameter. At first we put a cylinder of one-layer permalloy in the midpoint between the two coils with the cylinder's axis perpendicular to the direction of 3G field. The magnitude of the field in the center of the cylinder was measured to be 0.16G, $\sim$1/18 of the coils' field.

The cylinder of one-layer permalloy was then wrapped around the R5610A with its edge level to the R5610A's entrance window. The PMT and the cylinder were located in the position where the PMT's X orientation was along the axis of Helmholtz coils. Running the data taking program in B=0G and B=3G respectively during the quietest period of the day, 3 sets of data were acquired in each situation. After fitting the spectra, 3 peak channels (ADC channel) were obtained and averaged. The results in Fig.6 show that a cylinder of one-layer permalloy strip of 27$\mu$m thickness and 25mm width is able to shield the R5610A from a field of 3G with its gain loss less than 1\%. This has met the requirement of the energy resolution of the calorimeter.
\begin{center}
\includegraphics[width=8.6cm]{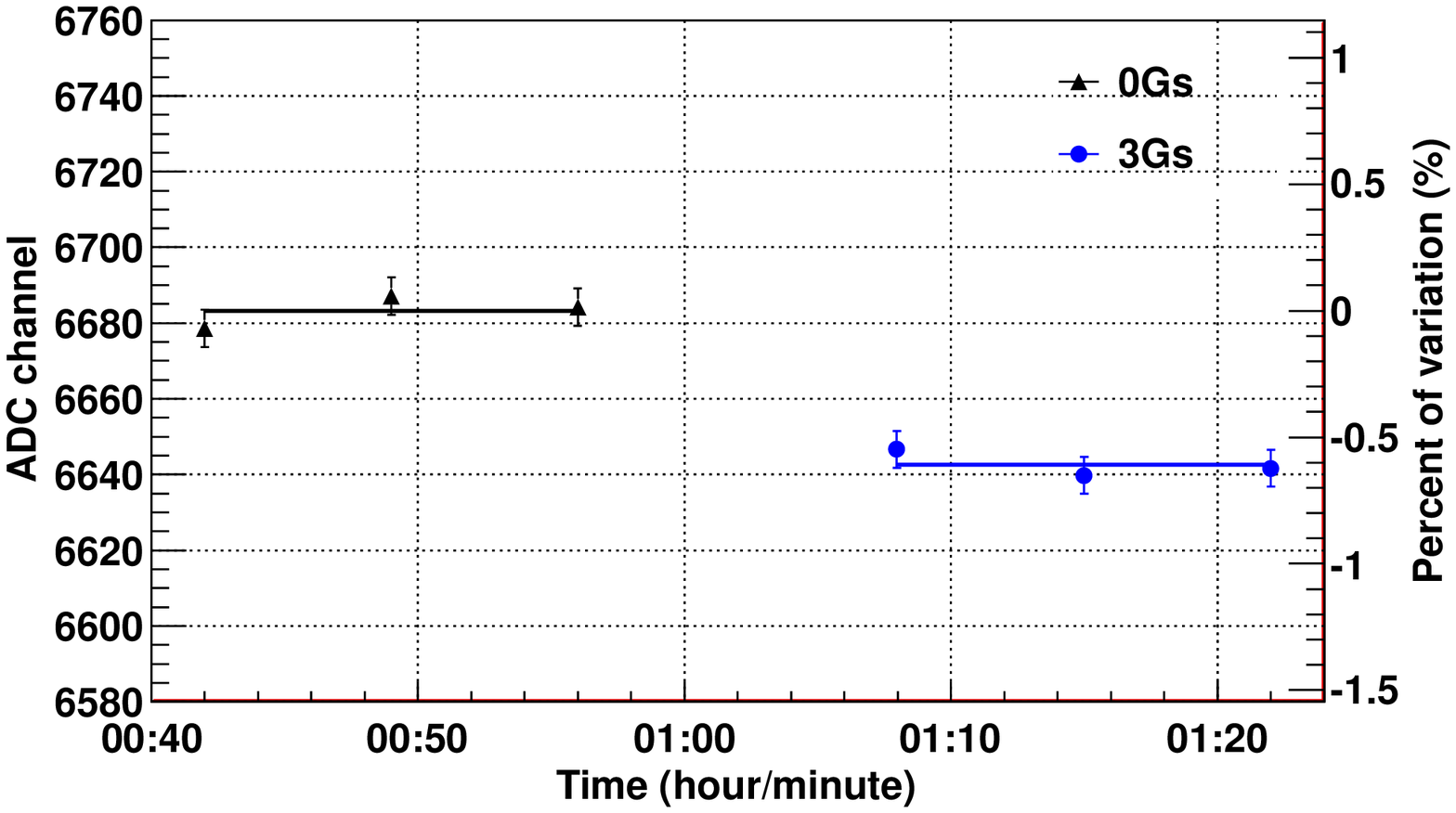}
\figcaption{\label{fig6}   The shield result of a cylinder of one-layer permalloy strip.}
\end{center}
\section{Conclusion}
The photomultiplier tube is very sensitive to external magnetic fields, because they seriously disturb the photo-electron paths defined by the dynode structure. Data shows that the gain of the R5610A loses ~53\% (+X orientation) and ~4\% (+Y orientation) when the magnetic magnitude is increasing from 0G to 3G. And along the Z orientation its gain has no visible change. A cylinder of one-layer permalloy is able to keep the gain loss of the R5610A to less than 1\% ($\sim$0.6\%) under 3G.\\

\acknowledgments{We would like to thank Professor Guangshun Huang for some extremely useful discussions. We are also very thankful to Mr. Grant Renny for his helpful English assistance. This work is supported by the 973 Program (Grant No. 2010CB833002), and the Strategic Priority Research Program on Space Science of the Chinese Academy of Science (Grant No. XDA04040202-4).}

\end{multicols}

\vspace{-1mm}
\centerline{\rule{80mm}{0.1pt}}
\vspace{2mm}

\begin{multicols}{2}

\end{multicols}

\clearpage

\end{document}